\documentclass[fleqn]{article}
\usepackage{graphicx}

\newcommand{\Kpi}{K^-\pi^+}
\newcommand{\KK}{K^-K^+}

\newcommand{\Dz}{D^0}
\newcommand{\Dzbar}{\bar{D^0}}
\newcommand{\D}{D^+}
\newcommand{\Dr}{D^*}

\newcommand{\DzKpi}{D^0 \rightarrow K^-\pi^+}
\newcommand{\DzKK}{D^0 \rightarrow K^-K^+}
\newcommand{\Dzpipi}{D^0 \rightarrow \pi^-\pi^+}

\newcommand{\DzwKpi}{D^0 \rightarrow K^+\pi^-}

\newcommand{\DKpi}{D \rightarrow K\pi}
\newcommand{\DKpp}{D \rightarrow K\pi\pi}
\newcommand{\DKppp}{D \rightarrow K\pi\pi\pi}
\newcommand{\DKK}{D \rightarrow KK}
\newcommand{\Dpipi}{D \rightarrow \pi\pi}

\newcommand{\ycp}{y_{CP}}
\newcommand{\xp}{x^\prime}
\newcommand{\yp}{y^\prime}
\newcommand{\tp}{t^\prime}

\newcommand{\fs}{\ {\mathrm fs}}
\newcommand{\fb}{{\mathrm fb}}
\newcommand{\ee}{\mathrm e^+e^-}

\begin{document}

\title{Recent Results on CP Lifetime Differences of Neutral D Mesons}

\author{Doris Yangsoo Kim
       \thanks{This work was supported by University of Illinois under
                contract  DE-FG02-91ER-40677 with the U.S. Department of
                Energy.} \\
        {\it Loomis Lab of Physics, University of Illinois,} \\
        {\it 1110 W. Green St., Urbana, IL. 61801, U.S.} }
\date{May 29, 2001}

\maketitle

\begin{abstract}
The mixing parameter $\ycp$ for the neutral $D$ system can be obtained by
comparing the lifetimes of CP eigenstate decays of the $D$ mesons.
Recent results on the lifetime differences from the Belle, CLEO, FOCUS(E831)
experiments are summarized in this article. The neutral $D$ decay modes 
analyzed by the experiments are:  The $\DzKpi$ mode with the assumption
that it is an equal mixture of CP even and odd 
eigenstates, and the CP even modes $\DzKK$ and $\Dzpipi$.
\end{abstract}

\section{Introduction}
Time dependent interactions of $\Dz$ and $\Dzbar$ decays can be coupled by the
off-diagonal terms in the Schr\"{o}dinger equation.
If CP is conserved in neutral $D$ decays, the eigenstates will be
described as ($\Dz \pm \Dzbar$)/$\sqrt{2}$ with mass and width as
$M \pm 1/2 \Delta M$, $\Gamma \mp 1/2 \Delta\Gamma$. Many calculations conclude 
that Standard Model contributions to $x (=\Delta M/\Gamma)$ and 
$y (= \Delta\Gamma/2\Gamma)$ parameters are very small, 
and new physics effects can manifest themselves as an 
unexpected large measurement in the $x$ parameter~\cite{MixTheory}.
Experimentally, the mixing parameters $x$ and $y$ are accessible either
by searching wrong sign decay of neutral $D$ mesons or by comparing
lifetimes of CP eigenstate decay channels.  In this article, we 
summarize recent efforts of various experiments who are exploring the
charm mixing sector using the second method. From now on, we assume that
the charge conjugate decay modes give the same results as the original decay 
modes.

For example, if we assume $\DzKpi$ decay or its charge conjugate mode
are advancing with equal amount of CP even and odd events, the width
of the decay channels is described
by $\Gamma(K\pi) \approx (\Gamma_1+\Gamma_2)/2$. Meanwhile, $\DzKK$ and
$\Dzpipi$ decay modes are CP even and their width is described by
$\Gamma_2$. Then $\ycp$ can be measured as:
\begin{equation}
\ycp = \frac{\tau(\DKpi)}{\tau(\DKK)} - 1 = 
       \frac{\tau(\DKpi)}{\tau(\Dpipi)} - 1
\end{equation}

In the following sessions, the analysis of the FOCUS experiment~\cite{Focus}
will be shown as well as the preliminary results from
the Belle~\cite{Belle} and the CLEO~\cite{Cleo2} experiments.
Implications of these measurements compared to the result of the
wrong sign $\Dz$ decay analysis done by the CLEO experiment~\cite{Cleo}
will be discussed at the end. 

\section{The FOCUS CP lifetime analysis}
The FOCUS (E831) collaboration consists of 100 physicists from USA, Italy,
Brazil, Mexico and Korea. The experiment is a successor to the Fermilab 
E687~\cite{E687} experiment,
and its main purpose is to study charm mesons and baryons produced by
photon beams of about 180 GeV at a fixed target spectrometer. 
During 1996-1997 run, 
the spectrometer collected more than 1 million reconstructed golden mode 
charm decays, $\DKpi$, $\DKpp$, and $\DKppp$. The FOCUS group used two
modes of neutral $D$ decays, $\DKpi$ and $\DKK$, to obtain the $\ycp$ 
parameter.  

\subsection{The FOCUS spectrometer performance}
The target material consists of 4 slabs of Be or BeO, depending on the data
run periods.
The segmented structure allows 62\% of charm particles produced in the target
material decaying in air, reducing the absorption correction necessary to the
lifetime analysis to a small amount. Two sets of silicon microstrip
vertex detectors are built to reconstruct vertices. 
The components of the first detector set interspace the target slabs, while the
second set is located downstream of the target. The combination of the 
relativistic momentum of charged particles and excellent vertex
detector performance produces an extremely good proper time resolution of
$30 \fs$ for neutral $D$ mesons, which corresponds to 8\% of its lifetime.
The good time resolution allows the FOCUS group to use a binned likelihood 
method free of deconvolution corrections to fit lifetime evolution of 
charm particles.  

The reconstruction algorithm of charm events is optimized to have a flat
acceptance over the fit variable, ``reduced proper time". 
The $D$ production and decay vertices are reconstructed by a candidate 
driven algorithm. A cut is imposed on the amount of detachment ($l/\sigma$)
between the production and the decay vertices to suppress backgrounds coming 
from light quark events. Reduced proper time is a variable used by fixed target 
experiments to measure lifetime of particles, defined as the difference between
the proper time and the amount of the detachment cut 
($\tp = t - N\sigma$). 

The FOCUS spectrometer has 3 stations of threshold Cerenkov detectors to 
identify charged particles. A continuous, likelihood-based identification
system is used. This algorithm gives the flexibility useful in 
assessing systematic errors due to backgrounds.
 
\subsection{Event Selection and Fitting}
The event selection is optimized to obtain a sample which has a flat efficiency
over the fit variable, $\tp$. First, neutral $D$ meson candidates are
required to pass basic section cuts of detachments ($l/\sigma > 5$) and
for tracks reconstructed as kaons, the Cerenkov hypothesis to be a kaon 
should be favored over the hypothesis to be a pion by a factor of 7.39.
Second, the candidate is either identified as a decay of $\Dr$ (tagged sample)
or required to pass more stringent particle identification cuts and kinematic
cuts (inclusive sample).  As a result, the FOCUS group obtained 119,738
$\DKpi$ candidates and 10,331 $\DKK$ candidates.

\begin{figure}[htb]
        \hspace*{-2.0cm}
        \includegraphics[height=2.in]{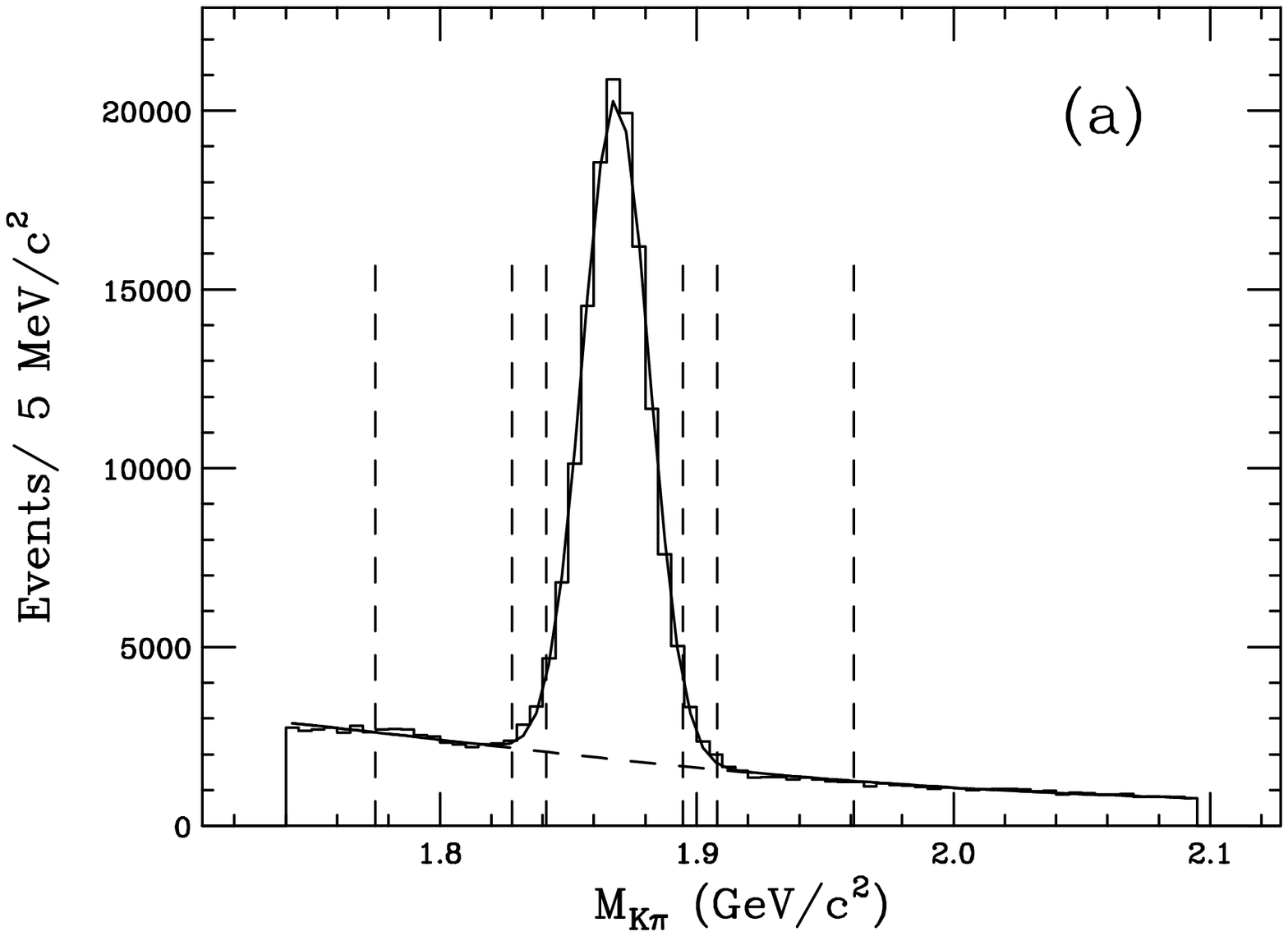}
        \includegraphics[height=2.in]{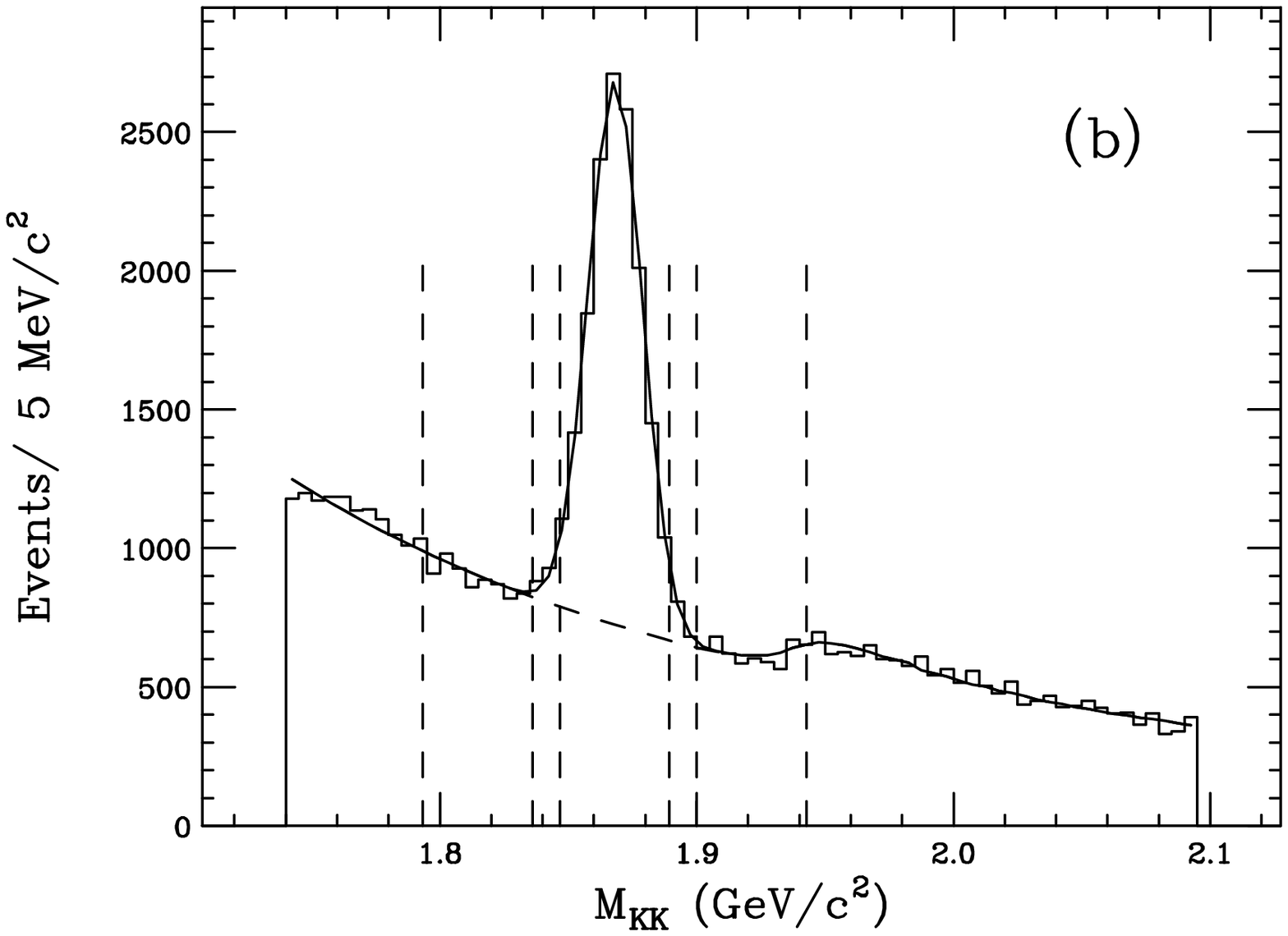}
        \begin{center}
         (c)\includegraphics[height=2.in]{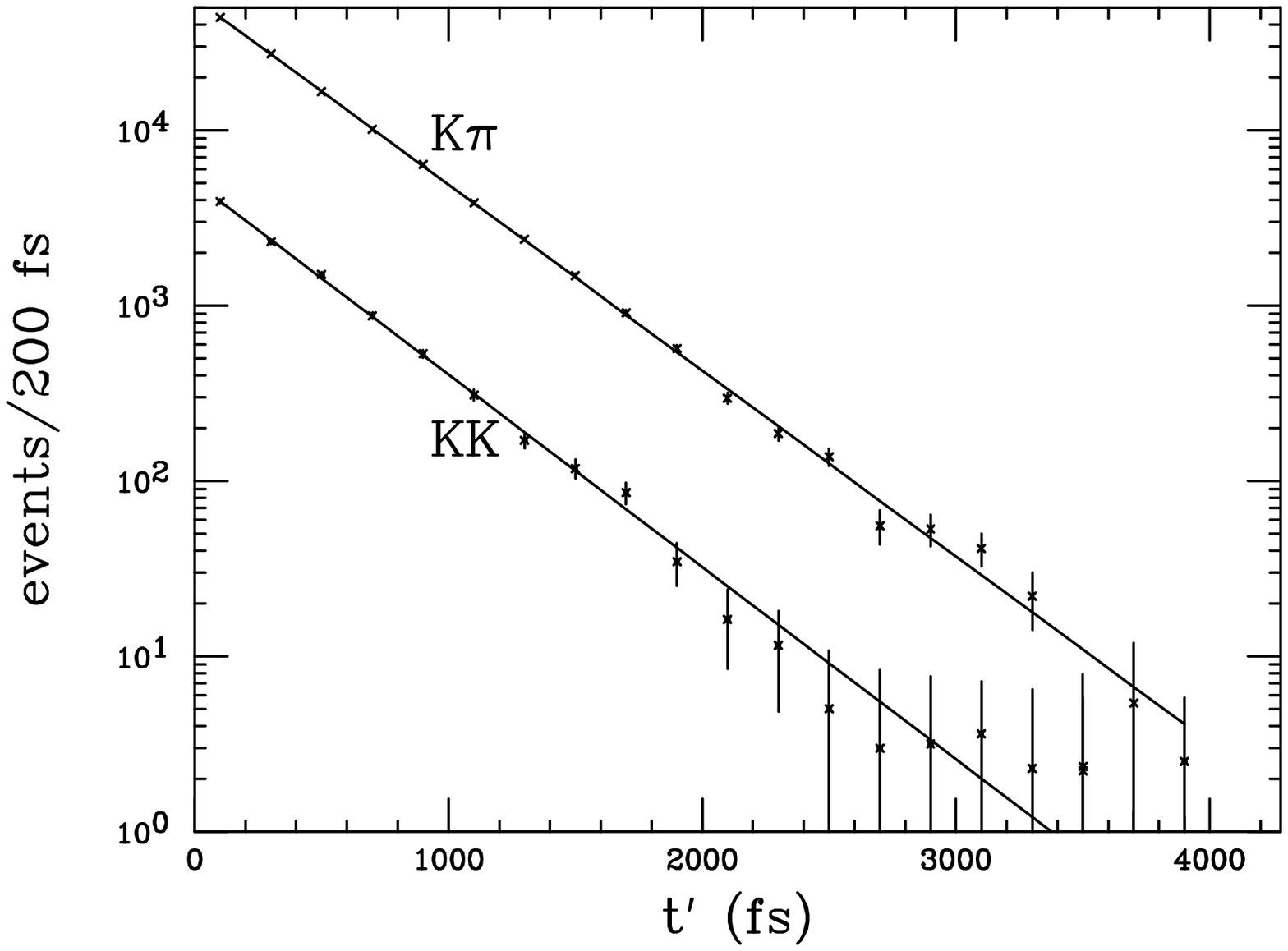}
        \end{center}
\caption{ (a) Reconstructed mass distribution of $\DzKpi$ and its
         conjugate candidates from the FOCUS experiment. 
         The yield is 119\,738 $\Kpi$ and $K^+\pi^-$ signal events;
          (b) That of $\DzKK$ candidates. The yield is
         10\,331 $\KK$ signal events. The vertical and dashed
         lines indicate the signal and sideband regions used for the
         lifetime and $\ycp$ fits;
          (c) Signal versus reduced proper time for the $\DKpi$ and $\DKK$ 
         events shown above. Each data point is background subtracted and
         includes the (very small) Monte Carlo correction;
         The plots shown here are reprinted from~\protect\cite{Focus}.}
\label{fig:focusplots}
\end{figure}

Histograms of 20 bins with $200 \fs$ width are used to fit the reduced proper
time of the $\DKpi$ and $\DKK$ candidates.
In each time bin, the amount of signal
is estimated as those of exponential decays adjusted by
the absorption/acceptance correction function $f(\tp)$ obtained from Monte
Carlo (MC) simulation. 
The amount of backgrounds is estimated from the sidebands of $D$ mass lineshape.
In case of the $\DKK$ sample, as shown in Figure~\ref{fig:focusplots} (b), 
a small amount of $\DKpi$ reflection background
is still lingering in the sideband area.
Since the time evolution of the reflection background in the
$\DKK$ sample is governed by the $\DKpi$ lifetime, lifetime fitting should
be done for both decay samples at the same time. The fit parameters are: $\DKpi$
lifetime ($\tau(\DKpi)$), the mixing parameter $\ycp$,
and normalization parameters of background events for the $\DKpi$ sample
($B_{K\pi}$) and for the $\DKK$ sample ($B_{KK}$).

\subsection{Results and Systematics}
Figure~\ref{fig:focusplots} (c) shows the reduced proper time evolution of the 
background subtracted $\DKpi$ and $\DKK$ candidates in the final fit result.
The lifetime and the lifetime differences of the neutral
$D$ mesons are fitted as follows:
\begin{eqnarray}
\ycp  = (3.42 \pm 1.39 \pm 0.74)\% & \\
\tau(\DKpi)  =  409.2 \pm 1.3 \fs &
\end{eqnarray}
The systematic error in $\ycp$ is obtained by varying selection cuts
and by changing lifetime fit options.  The systematic error on $\tau(\DKpi)$
is not shown, since inputs from other neutral $D$ decay channels, for example,
$\DKppp$, are needed. 
 
\section{The Belle $\Dz$ lifetime study}
The Belle collaboration presented a preliminary measurements
on lifetimes of $\Dz$, $\D$ and $D_S$ mesons at the 4th International 
Workshop on B physics and CP Violation (BCP4)~\cite{Belle}. 
The neutral charm mixing study is
shown as a part of their lifetime analysis package, based on the data set of
$11/\fb$ produced by an asymmetric $\ee$ collider, KEK-B, operating 
at 12 GeV $\Upsilon(4s)$.
The mixing study is performed with two neutral $D$ meson decay modes,
$\DzKpi$ and $\DzKK$.

\subsection{Event Selection}
To suppress the $b\bar{b}$ background smearing into the charm sample,
the momentum of $D$ candidates is asked to be more than 2.5 GeV in the
$\Upsilon(4s)$ frame. 
Identification of charged particles is determined by using information from
various detectors: a central drift chamber (CDC), 
a time of flight system (TOF), and an aerosol Cerenkov counter (ACC). 
$K/\pi$ separation of up to 3.5 GeV is achieved.
To enhance purity of the $D$ sample further, a decay angle cut is applied for
each mode. For signal events, the decay angle distribution ($\cos \theta_D$)
should be flat. The requirement of the $\Dr$ tag is optional. From the
selection, the Belle group found $90,601 \pm 387$ $\DKpi$ events and
 $7,451 \pm 118$ $\DKK$ events. 

Proper lifetimes of charm mesons are calculated from the distance between
the charm decay vertex and the production vertex
($t = l \cdot m(D)\,/\,p(D) \cdot c$). 
The decay vertex is reconstructed using all tracks from the charm candidate.
The resulting D flight path is extrapolated to the beam interaction region and
a production vertex is obtained. 

\subsection{Lifetime and $\ycp$ fit results}
An unbinned maximum likelihood fit is used to obtain the charm lifetime
and the mixing variable $\ycp$. 
The likelihood function for each event is described by a sum of following 
terms: The signal term described by an exponential of lifetime $\tau_{SIG}$ 
convoluted by a resolution function $R_{SIG}$, a term for background
events described by an exponential of lifetime $\tau_{BG}$ convoluted by
a resolution function $R_{BG}$, and another term for background events
which do not have a lifetime and convoluted by the $R_{BG}$.
For signals (backgrounds),
the resolution is $155\ (160)\fs$ for 56 (60)\% of events and 
$348\ (350)\fs$ for 44 (40)\% of events.

\begin{figure}[htb]
        \hspace*{-.5cm} 
        (a)\includegraphics[height=2.5in]{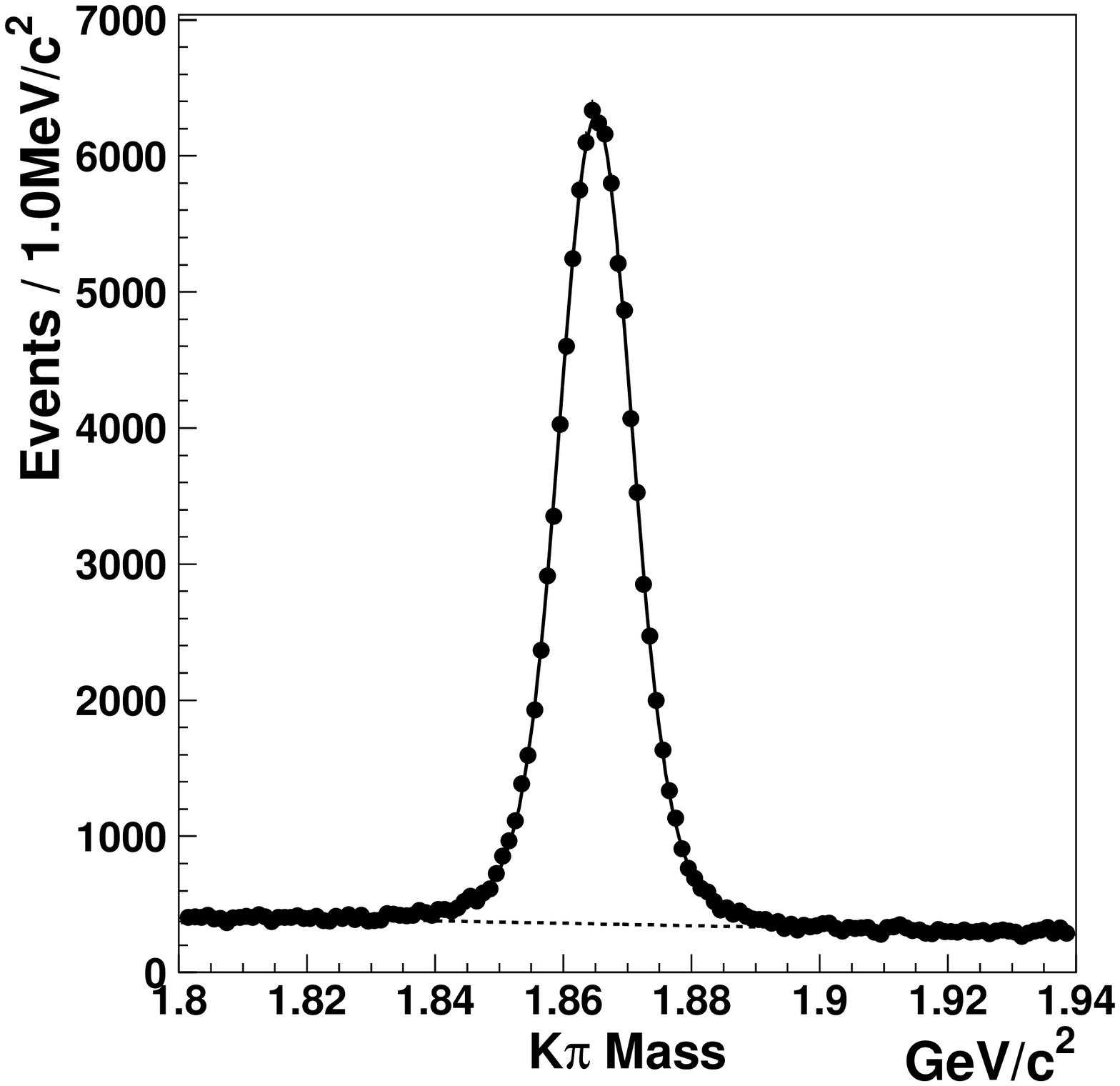}
        (b)\includegraphics[height=2.5in]{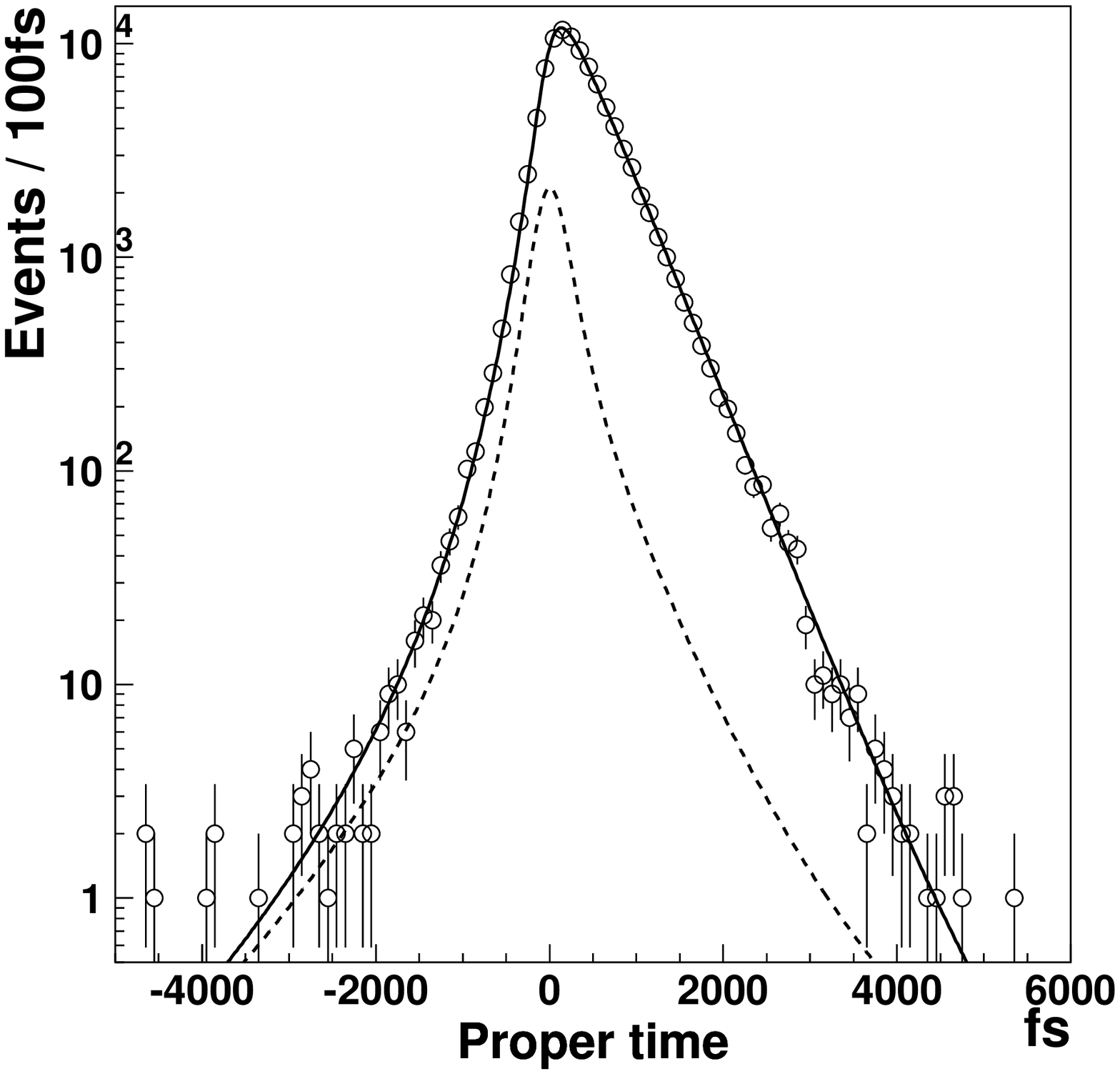}\\
        \hspace*{-.5cm}
        (c)\includegraphics[height=2.5in]{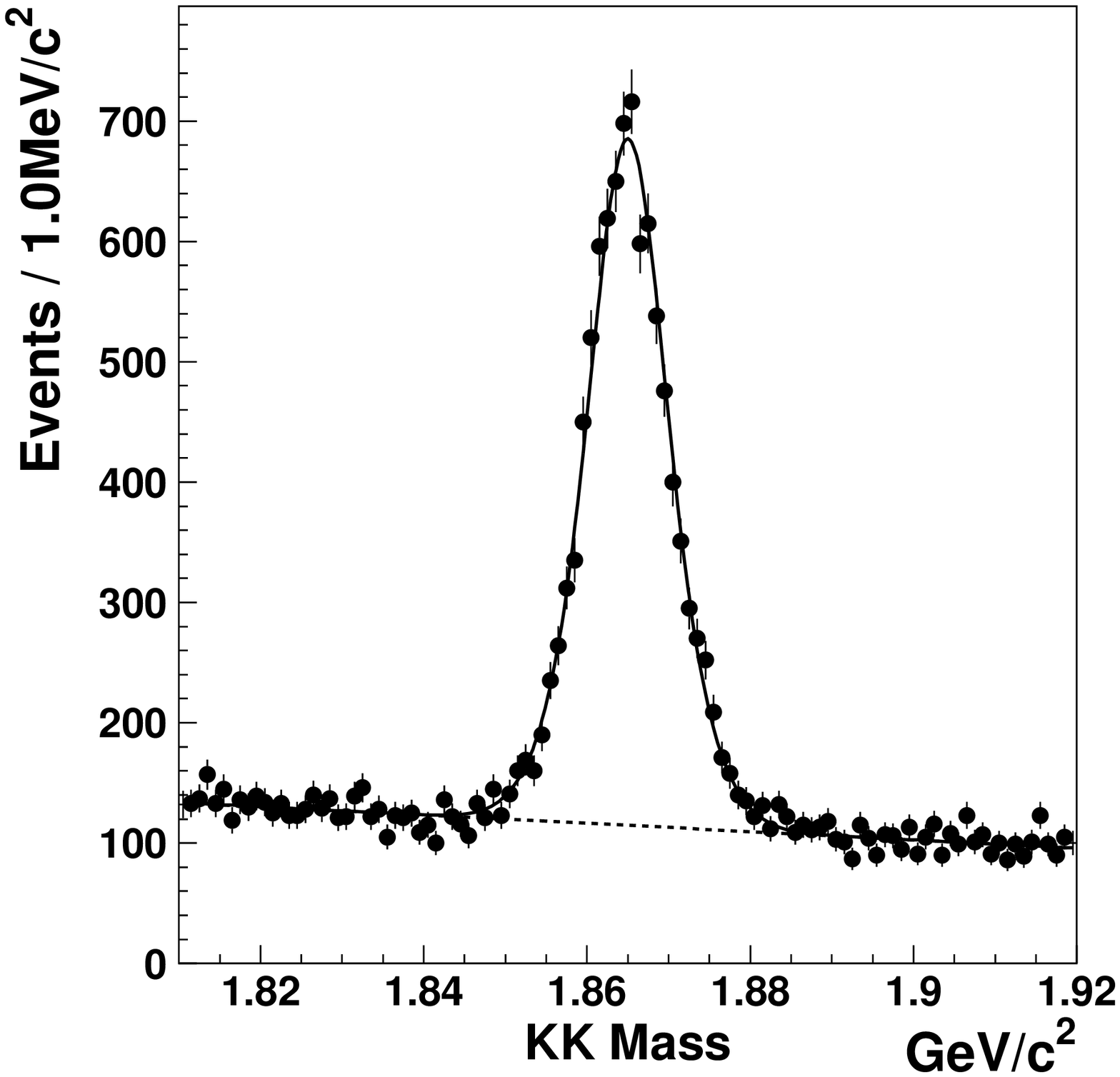}
        (d)\includegraphics[height=2.5in]{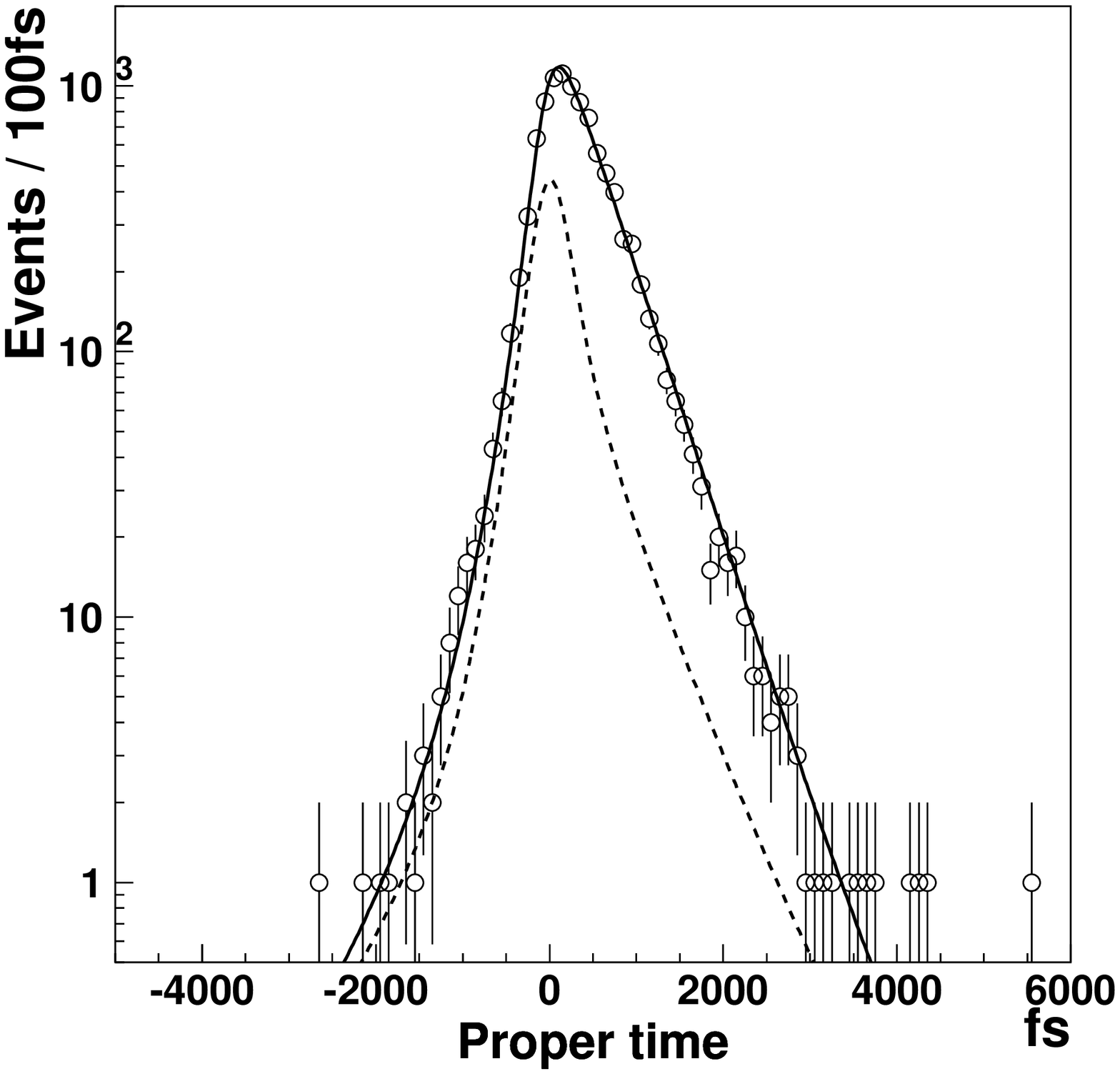}
\caption{ Preliminary results from the Belle experiment on neutral $D$
        lifetime differences~\protect\cite{Belle};
          (a) Reconstructed mass distribution of $\DzKpi$ and its
         conjugate candidates.
         The yield is $90,601 \pm 387$ $\Kpi$ and $K^+\pi^-$ signal events;
          (b) The proper time distribution of the $\DKpi$ sample.
         The dashed line is the estimated background in the sample;
          (c) Reconstructed mass distribution of $\DzKK$ candidates.
         The yield is $7,451 \pm 118$ $\KK$ signal events;
          (d) The proper time distribution of the $\DKK$ sample.
    The dashed line is the estimated background in the sample.}
\label{fig:belleplots}
\end{figure}

A combined lifetime fit using both $\DKpi$ and $\DKK$ distributions is 
performed and the following lifetime parameters are obtained: 
\begin{eqnarray}
\lefteqn{ \ycp  = 1.16^{+1.67}_{-1.65}\, \% }\\
\lefteqn{ \tau(\DKpi)  =  414.5 \pm 1.7 \fs} 
\end{eqnarray}
The errors shown above are statistical only.  A detailed study on systematic
errors is currently going on.
The numbers are consistent with separately measured neutral 
$D$ lifetimes; $\tau(\DKpi) = 414.5 \pm 1.7 \fs$, 
$\tau(\DKK) = 409.8 \pm 6.3 \fs$
The Belle measurements on the $\Dz$ lifetime and the mixing parameter $\ycp$ are
consistent with the ones from other experiments. 
Figure~\ref{fig:belleplots} shows the mass and lifetime distributions used
to extract the fit results.  

\section{The CLEO $\Dz$ lifetime study}
At the BCP4 conference, the CLEO collaboration summarized their recent 
efforts on measurements of $\Dz - \Dzbar$ mixing, CP violation in $\Dz$ decays,
and $D^{*+}$ width~\cite{Cleo2}. They showed the preliminary analysis
on $\Dz$ lifetime using $\DKpi$, $\DKK$, and $\Dpipi$ decay channels,
based on a data set of $9.0/\fb$ produced by a symmetric $\ee$ collider, CESR,
operating at 10.6 GeV ($\Upsilon(4s)$).   

\subsection{Event Selection}
The kinematic characteristics of charm particles created in the CLEO detector
are similar to those created in the Belle. The $\Dr$ tag is required to
select neutral $D$ events, exploiting the excellent CLEO resolution
on the mass difference between $\Dr$ and $\Dz$ ($\sigma_Q = 190 \pm 2 KeV$).
The CLEO group selected $20,272 \pm 178$ $\DKpi$,
$2,463 \pm 65$ $\DKK$ and $930 \pm 37$ $\Dpipi$ events.  

\subsection{Lifetime and $\ycp$ fit results}
The CLEO analysis is also performed using unbinned maximum likelihood fits
to proper time distributions of three $\Dz$ decay channel candidates. 
The fitting method is optimized for finding lifetime 
differences, not for finding lifetime of neutral $D$ mesons. The resolution
function of proper time in the likelihood is described by three Gaussians:
One Gaussian for events with correctly measured proper time and two Gaussians
for events with mismeasured proper time. For most events, the resolution
is about 35\% of the neutral $D$ lifetime. The following $\tau(\Dz)$
values are obtained for each decay channel: $\tau(\DzKpi) = 404.6 \pm
3.6 \fs$,  $\tau(\DzKK) = 411 \pm 12 \fs$, and $\tau(\Dzpipi) = 401 \pm
17 \fs$. The $\tau(\Dz)$ values are highly correlated
to the $\Dz$ mass constrained in the fit procedure. This fit technique
is not used in the dedicated $D$ lifetime analysis published by the
CLEO group~\cite{Cleo3}. Mixing parameter values
are calculated as differences in lifetimes among three decay channels: 
\begin{eqnarray}
\lefteqn{\ycp(KK)  = (-1.9 \pm 2.9 \pm 1.6)\% }  \\
\lefteqn{\ycp(\pi\pi)  = (0.5 \pm 4.3 \pm 1.8)\% }
\end{eqnarray}
\begin{equation}
\ycp(\mathrm{combined})  = (-1.1 \pm 2.5 \pm 1.4)\%
\end{equation}
This preliminary result is consistent with the measurements from the other
experiments. Dominant systematic errors are coming from statistical uncertainty
in MC lifetime study ($9 \fs$), background description ($8 \fs$), proper
time resolution model ($5 \fs$), and fit procedure ($5 \fs$).

\section{Conclusion}
The most recent five results on $\Dz$ lifetime are shown 
on Table~\ref{lifetime}. The measurements obtained by
the FOCUS and the Belle experiments from the lifetime difference analysis
are comparable to those from the dedicated $\Dz$ lifetime analysis.
\begin{table} \label{lifetime}
\begin{center}
\begin{tabular}{ccc} \hline\hline
PDG2K                    & $412.6 \pm 2.8$~\cite{PDG2K} & \\ \hline 
                         &                              & \\ \hline
E791                     & $413 \pm 3 \pm 4$~\cite{E791}& included in PDG2K \\
CLEO                     & $408.5 \pm 4.1 \pm 3.5$~\cite{Cleo3}  
                                                        & included in PDG2K \\
FOCUS (E831)             & $409.2 \pm 1.3 \pm x^*$~\cite{Focus}  & \\ 
SELEX (E781)             & $407.9 \pm 6.0 \pm 4.3$~\cite{Selex}  & \\ 
Belle preliminary        & $414.5 \pm 1.7 \pm x^*$~\cite{Belle}  & \\ \hline
                         &                          & \\ \hline
Average of               & $411.1 \pm 1.0$          & \\ 
recent values            & $\chi^2 = 6.7$ for $dof=4$ & \\ \hline \hline
\end{tabular}
\end{center}
\caption{The recent results on $\Dz$ lifetime from various experiments
are shown in chronological order. 
The average in the table is obtained from the most recent 
five measurements. The numbers are consistent between one another.
The E791 and CLEO values are already included in the PDG 2K
edition~\cite{PDG2K}. $x^*$ denotes
that corresponding systematic errors are not shown by the experiments yet.} 
\end{table}

Table~\ref{ycp} shows compilation of $\ycp$ values obtained by 
various experiments. The numbers are comparable to one another, producing
an average of $(1.8 \pm 1.0)\%$ with $\chi^2/dof = 2.3/3$.
On the other hands, the CLEO experiment published an impressive paper
on the neutral
$D$ mixing parameters, $\xp$ and $\yp$ by analyzing lifetime evolution
of wrong sign decays of neutral $D$ mesons ($\DzwKpi$)~\cite{Cleo}. 
On the surface, the average $\ycp$ value is comparable to the CLEO 
wrong sign results, $-5.8\% < \yp < 1\%$ at $95\% CL$. 
But one has to be careful when combining $\ycp$ and $\yp$ together. 
The $\yp$ and $\ycp$ parameters are differed by a strong angle $\cos\delta$
originated from final state interactions. The strong angle has been
considered to be small, but a recent theoretical paper suggested possibilities
of not so small $\cos\delta$~\cite{Falk}. 
The results on two $y$ parameters have comparable errors
but opposite signs, which prompted interesting discussions and
speculations on neutral charm mixing~\cite{BGLNP}.
This situation implies that mixing measurements in charm sector 
are challenging studies. We are eagerly waiting for up-to-date inputs from 
B-factories and next generation experiments to further enhance
understanding in ever fascinating mixing phenomena. 

\begin{table} \label{ycp}
\begin{center}
\begin{tabular}{cc} \hline\hline
E791               & $(0.8 \pm 2.9 \pm 1.)\%$~\cite{E791}    \\
FOCUS (E831)       & $(3.42 \pm 1.39 \pm 0.74)\%$~\cite{Focus} \\ 
Belle preliminary  & $(1.16^{+1.67}_{-1.65})\%$~\cite{Belle} \\
CLEO  preliminary  & $(-1.1 \pm 2.5 \pm 1.4)\%$~\cite{Cleo2} \\ \hline
                   &                                         \\ \hline
Average of $\ycp$  & $(1.8 \pm 1.0)\%$                       \\ 
                   & $\chi^2 = 2.3$ for $dof=3$              \\ \hline \hline
\end{tabular}
\end{center}
\caption{The recent results on the $\Dz$ lifetime difference $\ycp$ obtained
by various experiments via $\DKpi$, $\DKK$, and $\Dpipi$ channels are shown
in chronological order.
The numbers are consistent between one another. One has to be careful
if $\ycp$ values should be compared with $\yp$ parameter measurements.}
\end{table}

\begin{figure}[htb]
        \vspace*{-2.0cm}
        \hspace*{0.5cm}
        (a)\includegraphics[height=2.3in]{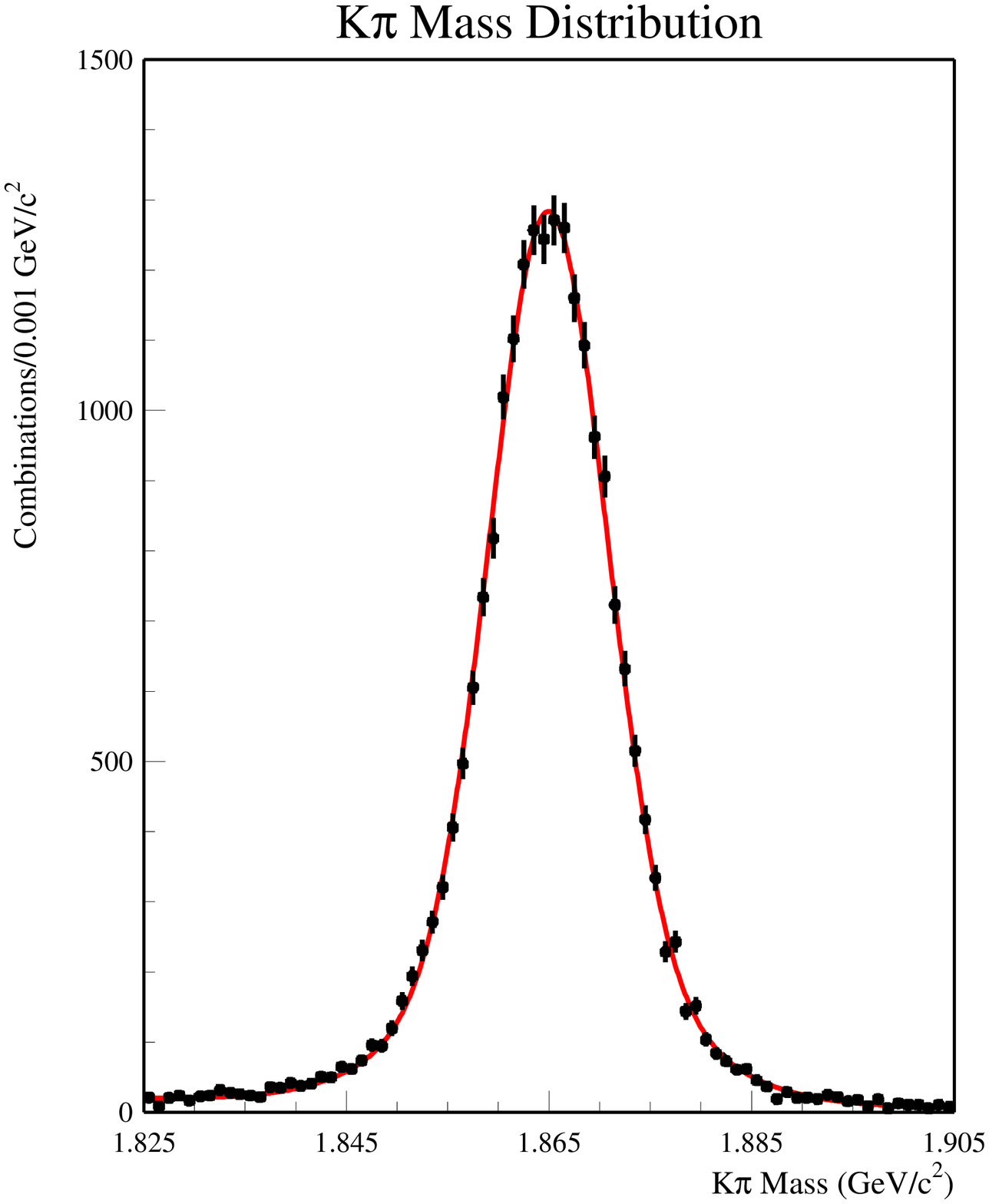}
        (b)\includegraphics[height=2.3in]{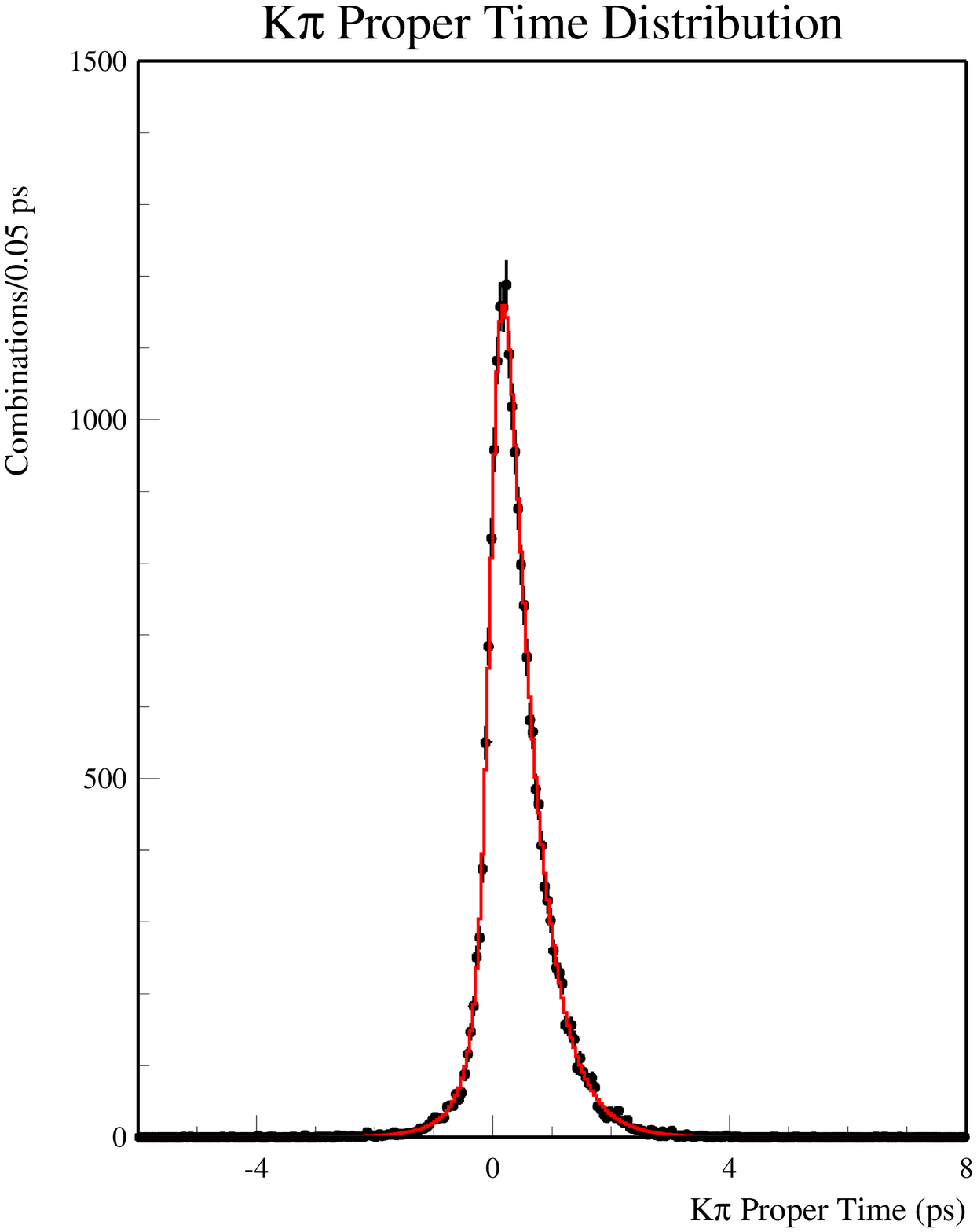}\\
        \hspace*{0.5cm}
        (c)\includegraphics[height=2.3in]{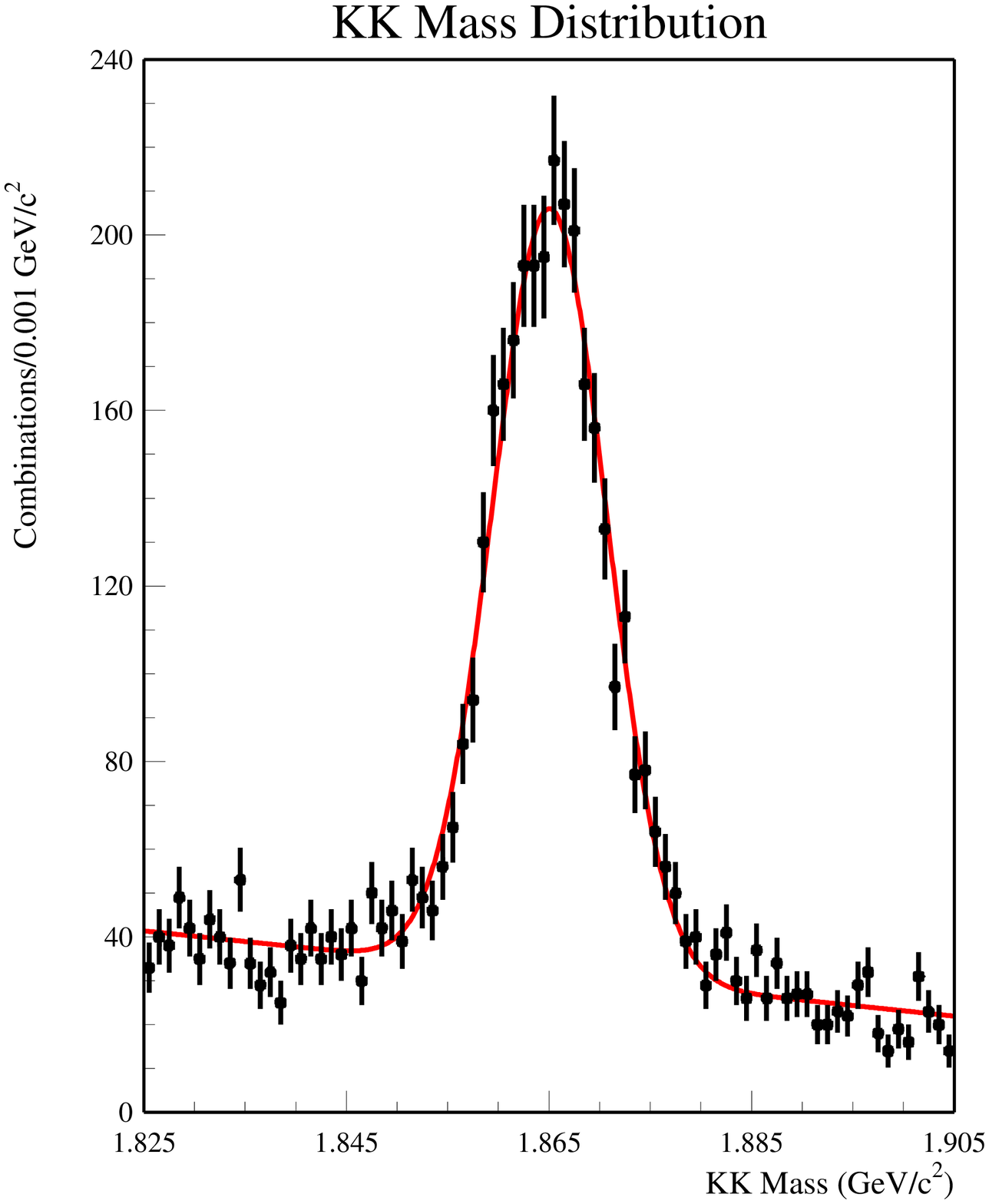}
        (d)\includegraphics[height=2.3in]{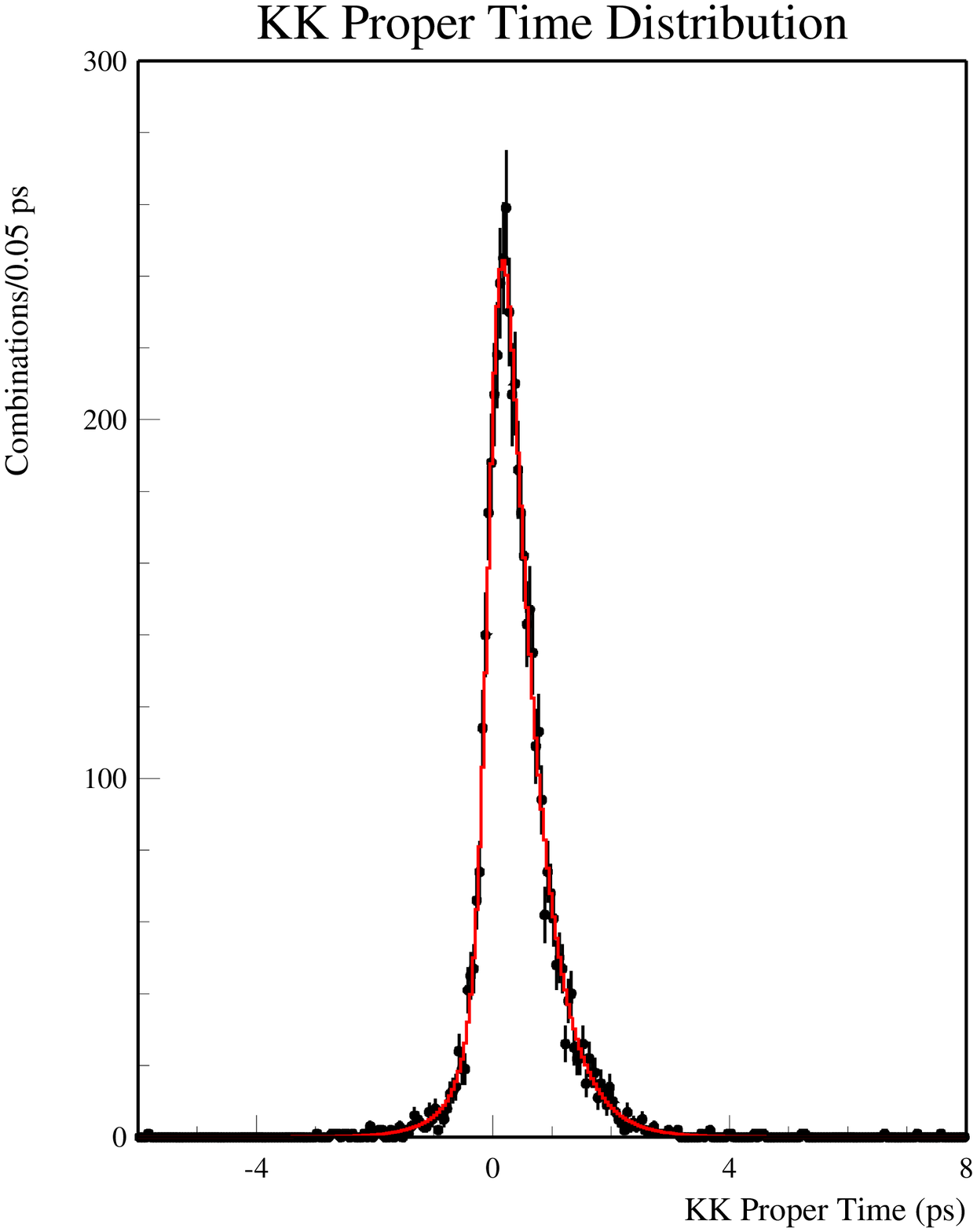}\\
        \hspace*{0.5cm}
        (e)\includegraphics[height=2.3in]{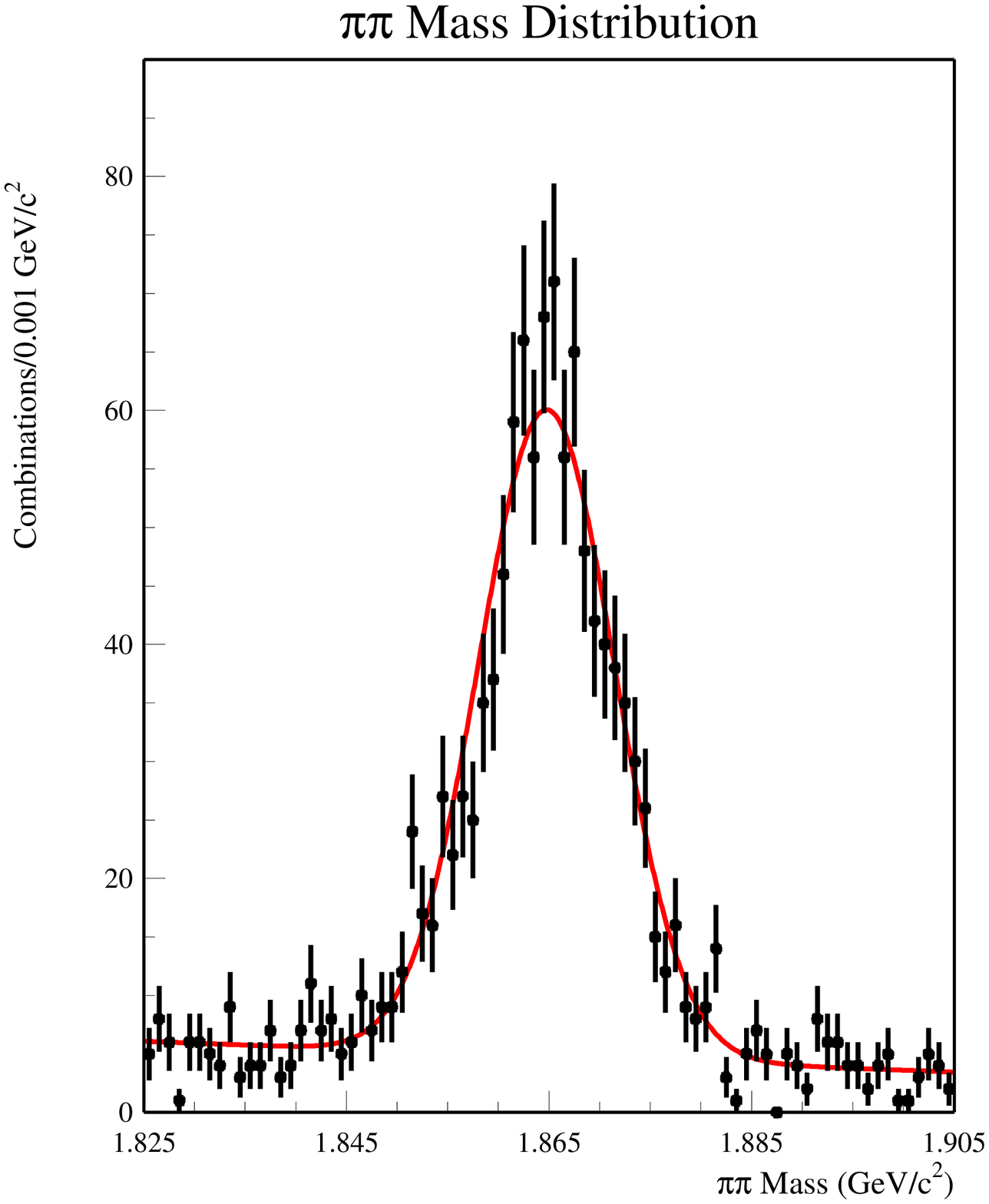}
        (f)\includegraphics[height=2.3in]{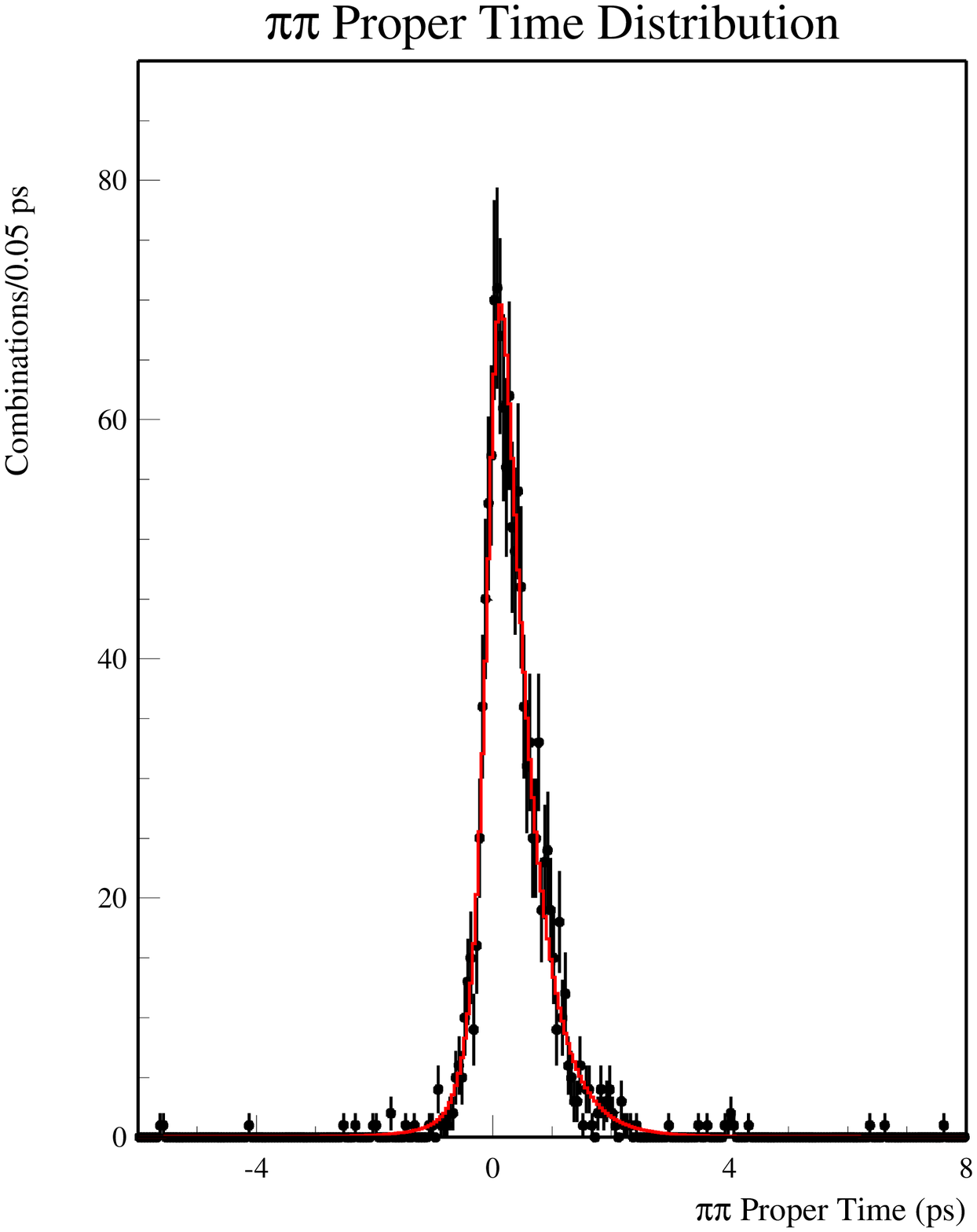}
\caption{ Preliminary results from the CLEO experiment on neutral $D$
        lifetime differences~\protect\cite{Cleo2};
         (a) Reconstructed mass distribution of $\DzKpi$ and its
        conjugate candidates.
        The yield is $20,272 \pm 178$ $\Kpi$ and $K^+\pi^-$ signal events;
         (b) The proper time distribution of the $\DKpi$ sample;
         (c) Reconstructed mass distribution of $\DzKK$ candidates.
        The yield is $2,463 \pm 65$ $\KK$ signal events;
         (d) The proper time distribution of the $\DKK$ sample;
         (e) Reconstructed mass distribution of $\Dzpipi$ candidates.
        The yield is $930 \pm 37$ $\KK$ signal events;
         (f) The proper time distribution of the $\Dpipi$ sample}
\label{fig:cleoplots}
\end{figure}                                                                     
\end{document}